\documentclass[runningheads]{llncs}
\usepackage[hidelinks]{hyperref} % hidelinks to remove url bounding box
\usepackage[T1]{fontenc}
\usepackage{graphicx}
\usepackage{hyperref}
\usepackage{cite}

\usepackage{xspace}
\usepackage{amsmath}
\usepackage{subfig}
\usepackage{orcidlink}
\usepackage{amsfonts}

\usepackage{enumitem}
\setlist{nosep}

\usepackage{xcolor}
\usepackage{colortbl}
\definecolor{lightgray}{gray}{0.9}

\makeatletter
\DeclareRobustCommand\onedot{\futurelet\@let@token\@onedot}
\def\@onedot{\ifx\@let@token.\else.\null\fi\xspace}

\def\etal{\emph{et al}\onedot}
\makeatother

\begin{document}
\title{DanceDuo: Bridging Human Movement and AI Choreography}
\titlerunning{DanceDuo: Bridging Human Movement and AI Choreography}
% If the paper title is too long for the running head, you can set
% an abbreviated paper title here

% TODO FINAL: Replace with your author list. 
% Include the authors' OCRID for the camera-ready version, if at all possible.
\author{
Gia-Cat Bui-Le\inst{1,2}\orcidlink{0009-0006-9936-3814} \and
Tuong-Vy Truong-Thuy\inst{1,2}\orcidlink{0009-0008-2541-576X} \and
Hai-Dang Nguyen\inst{1,2}\orcidlink{0000-0003-0888-8908} \and
Trung-Nghia Le\thanks{Corresponding author.}\inst{1,2}\orcidlink{0000-0002-7363-2610}}

% TODO FINAL: Replace with an abbreviated list of authors.
\authorrunning{G.-C. Bui-Le et al.}
% First names are abbreviated in the running head.
% If there are more than two authors, 'et al.' is used.

% TODO FINAL: Replace with your institution list.
\institute{University of Science, VNU-HCM, Ho Chi Minh City, Vietnam \and
Vietnam National University, Ho Chi Minh City, Vietnam
\email{\{blgcat20,tttvy20\}@apcs.fitus.edu.vn, nhdang@selab.hcmus.edu.vn, ltnghia@fit.hcmus.edu.vn}}

\maketitle              % typeset the header of the contribution
\begin{abstract}
In recent years, advancements in deep learning and generative models have revolutionized music-driven dance generation. This paper introduces a novel platform, namely DanceDuo, leveraging diffusion models to generate AI-choreographed dance sequences synchronized with a variety of music genres, to encourage dancing practice. The system allows users to interact with AI by selecting music tracks, humanoid models, and importing personal dance videos for comparison, fostering a rich and engaging user experience. DanceDuo not only offers dance generation but also integrates human pose estimation models to provide users with insightful comparisons of their own performances with AI-generated sequences. We conducted a comprehensive user study, revealing that users found the interface intuitive, with particular praise for the dance comparison feature. Our DanceDuo contributes significantly to the integration of AI in dance choreography, offering novel avenues for both recreational and professional applications. 

\end{abstract}

\section{Introduction}

In recent years, the advent of deep learning has driven rapid advancements in generative modeling, significantly impacting multimedia fields \cite{brown2020language, li2022diffusion, karras2019style, saharia2022palette, li2022srdiff, ho2022video, yang2023diffusion, zeng2023ipdreamer, poole2022dreamfusion}. Data synthesis using generative models is crucial for a range of applications \cite{zhu2024distribution, yang2024editworld, yang2024mastering, ramesh2022hierarchical, singer2022make, blau2022threat, yoon2021adversarial}. 
%
%
% Generative models possess the remarkable ability to learn complex patterns from existing datasets, enabling the production of new, synthetic data that closely resembles the original. This capability is crucial for a range of applications, such as data augmentation \cite{zhu2024distribution, yang2024editworld}, creative content generation \cite{yang2024mastering, ramesh2022hierarchical, singer2022make}, and enhancing the robustness of machine learning models \cite{blau2022threat, yoon2021adversarial}. The significance of generative modeling lies in its potential to automate traditionally manual processes, thereby improving efficiency and facilitating novel avenues for exploration and application.
%
One of the most captivating applications of generative modeling is music-to-dance generation \cite{mypaper, tseng2023edge, li2024lodge, siyao2022bailando}, which involves translating musical elements such as tempo, rhythm, and melody into coherent and expressive dance sequences. Recent advancements in deep learning architectures and generative models have made it possible to achieve this translation, enabling the creation of synchronized and harmonious dance movements. However, existing solutions \cite{li2021ai, yang2023longdancediff, tseng2023edge, li2024lodge, siyao2022bailando} often lack an interactive and engaging experience, limiting their potential to inspire creativity and facilitate learning.

To address this gap, we present DanceDuo, an novel application designed to harness the power of generative models for in-the-wild music-to-dance generation. By leveraging cutting-edge diffusion models \cite{mypaper}, DanceDuo provides users with an interactive platform to create dance sequences synchronized with a wide range of musical tracks. The application also features a diverse selection of humanoid models, allowing users to visualize AI-choreographed movements in rich detail. Beyond simply showcasing generated dances, DanceDuo also allows for the importation of personal dance videos, enabling users to compare their own movements with those generated by the AI by integrating human pose estimation models, thus facilitating an engaging learning experience that inspires users to explore their dancing potential. We also design a  scoring formula specifically that fosters engagement and motivation for all users, regardless of their dance proficiency, to encourage people to do excises via dancing. Moreover, this interaction helps in promoting creativity, as users can reflect on their own dance styles in conjunction with AI-generated sequences.

To evaluate the efficacy and user experience of our DanceDuo, we conducted a comprehensive user study. The study utilized an open-ended survey format to gather qualitative feedback on various aspects of the application, including usability, interface design, and feature effectiveness. Findings revealed that participants generally perceived DanceDuo as intuitive and engaging, especially praising the dance comparison feature for its interactive and motivational qualities.

The main contributions of this work are as follows:

\begin{itemize}
    \item Integration of a music-to-dance model facilitates the creation of an interactive platform that enhances user engagement and creativity in dance.
    \item Exploration of interactive features, such as the ability to compare a user's dance performance with AI-generated dances, provides valuable insights for skill improvement.
    \item Scoring formula is designed to foster engagement and motivation for all users, regardless of their dance proficiency.
    \item Comprehensive quantitative and qualitative evidences showcase the capabilities of DanceDuo.
\end{itemize}

\section{Related Work}

\subsection{Music-Driven Dance Generation}

%which selected pre-defined motion segments from a database and arranged them to align with music

%These networks utilize music and prior dance sequences as inputs to predict subsequent dance movements autoregressively. 

Early research in music-driven dance generation primarily employed retrieval-based techniques \cite{fan2011example, shiratori2006dancing}. However, recent methodologies have shifted towards motion synthesis approaches, leveraging advanced network architectures such as CNNs \cite{kritsis2022danceconv, holden2016deep}, RNNs \cite{huang2020dance, tang2018dance}, GANs \cite{sun2020deepdance, lee2019dancing}, and Transformers \cite{li2020learning, li2021ai, li2022danceformer}. Notably, the VQ-VAE framework \cite{van2017neural} has been employed to generate temporally coherent dance sequences \cite{siyao2022bailando, zhuang2023gtn}. Recently, diffusion models have shown significant promise in generating high-quality images, videos, and motion sequences \cite{rombach2022high, ruiz2023dreambooth, ho2022imagen, kong2020diffwave, singer2022make, kim2023flame, dabral2023mofusion}, with the EDGE framework demonstrating a music-conditioned motion denoising approach \cite{tseng2023edge}.%The EDGE framework \cite{tseng2023edge} demonstrated the use of diffusion models in dance generation by framing it as a music-conditioned motion denoising problem. Utilizing a transformer decoder for music conditioning, EDGE generated overlapping dance segments during the denoising process and ensured their consistency through diffusion inpainting \cite{lugmayr2022repaint}.

\subsection{Dance Interactive Systems}

The integration of digital tools in dance creation has expanded significantly with technological advancements. Notable examples include Merce Cunningham's ``Biped'' \cite{abouaf1999biped}, which combined computer-captured dance movements with hand-drawn graphics to create animated and abstract dance characters. Burton \etal \cite{burton2016laban}'s work on Laban Movement Analysis (LMA) highlights the importance of understanding movement dynamics in enhancing the relationship between dancers and technology.

Several studies have developed interactive environments that support the creation, practice, and performance of choreography. For instance, a virtual reality-based tele-immersive environments \cite{chan2010virtual} allows dancers to engage in immersive experiences, while interactive augmented reality systems enhance live performances \cite{brockhoeft2016interactive}. Additionally, Sheppard \etal \cite{sheppard2008advancing} introduced an application enabling multiple participants to interact regardless of physical distance, leading to the concept of tele-immersive dance (TED), which fosters a highly interactive collaborative environment.

Within the realm of human-computer interaction (HCI), various technologies have been employed to capture and analyze dance movements. These include vision-based motion capture systems \cite{aristidou2015folk, tsampounaris2016exploring}, depth cameras like Microsoft Kinect \cite{alexiadis2011evaluating, kyan2015approach}, and inertial motion sensors \cite{kitsikidis2014multi}. Moreover, a significant body of research has focused on translating movement assessment results into effective feedback \cite{trajkova2016ballet, trajkova2018takes}. By integrating these technological advancements, the field of dance continues to evolve, offering new opportunities for creativity and expression.

\subsection{3D Pose Estimation}

%Estimating the human body pose in 3D from a single image or video presents a significant challenge in computer vision. Existing methods can generally be categorized into two main approaches: multi-stage and single-stage techniques.

Estimating human body pose in3D from a single image or video is a significant challenge in computer vision. Methods can be categorized into multi-stage and single-stage techniques. Multi-stage methods involve detecting individuals and then estimating body pose parameters \cite{pavlakos2019texturepose, kanazawa2018end}. Recent advancements leverage supervisory signals to improve geometric and dynamic consistency \cite{rempe2021humor, yuan2022glamr}. Alternative single-stage solutions \cite{benzine2020pandanet, mehta2018single} directly estimate body joint positions and group them into individuals, such as the one-stage approach proposed by ROMP \cite{sun2021monocular}.

%In multi-stage methods, the conventional framework typically involves two stages: first, the detection of individuals, and subsequently, the estimation of body pose parameters for each person \cite{pavlakos2019texturepose, kolotouros2019learning, kanazawa2018end}. Recent advancements in this area have focused on leveraging various supervisory signals \cite{rempe2021humor, yuan2022glamr, kocabas2020vibe} to improve geometric and dynamic consistency in body pose estimation.

%such as temporal coherence, contour alignment, self-contact, ground constraints, and global human trajectory analysis. 

%Alternatively, one-stage solutions offer a more streamlined approach by reasoning about all individuals in a single forward pass. In this method, the model \cite{benzine2020pandanet, mehta2018single} directly estimates the positions of all body joints and subsequently group them into individuals. ROMP \cite{sun2021monocular} further advances the one-stage paradigm by extending the end-to-end process beyond mere joint estimation.

%\subsection{Motion Capture}

\section{DanceDuo Design}

\subsection{Overview}

DanceDuo is a versatile platform that leverages the power of diffusion models to deliver an interactive experience centered around AI-generated dance movements. By harnessing the power of these models, DanceDuo generates real-time dance sequences that synchronize seamlessly with a diverse selection of music tracks across various genres. The platform encompass a wide range of dance styles, including Break, Lock, LA Style Hip-hop, Waack, Street Jazz, Pop, Middle Hip-hop, House, Krump, and Ballet Jazz. These categories align with the AIST++ dataset \cite{li2021ai}, which is widely recognized for its applications in music-to-dance generation tasks.

\begin{figure}[t!]
\centering
\subfloat[Boy humanoid model.]{\includegraphics[width=0.4\textwidth]{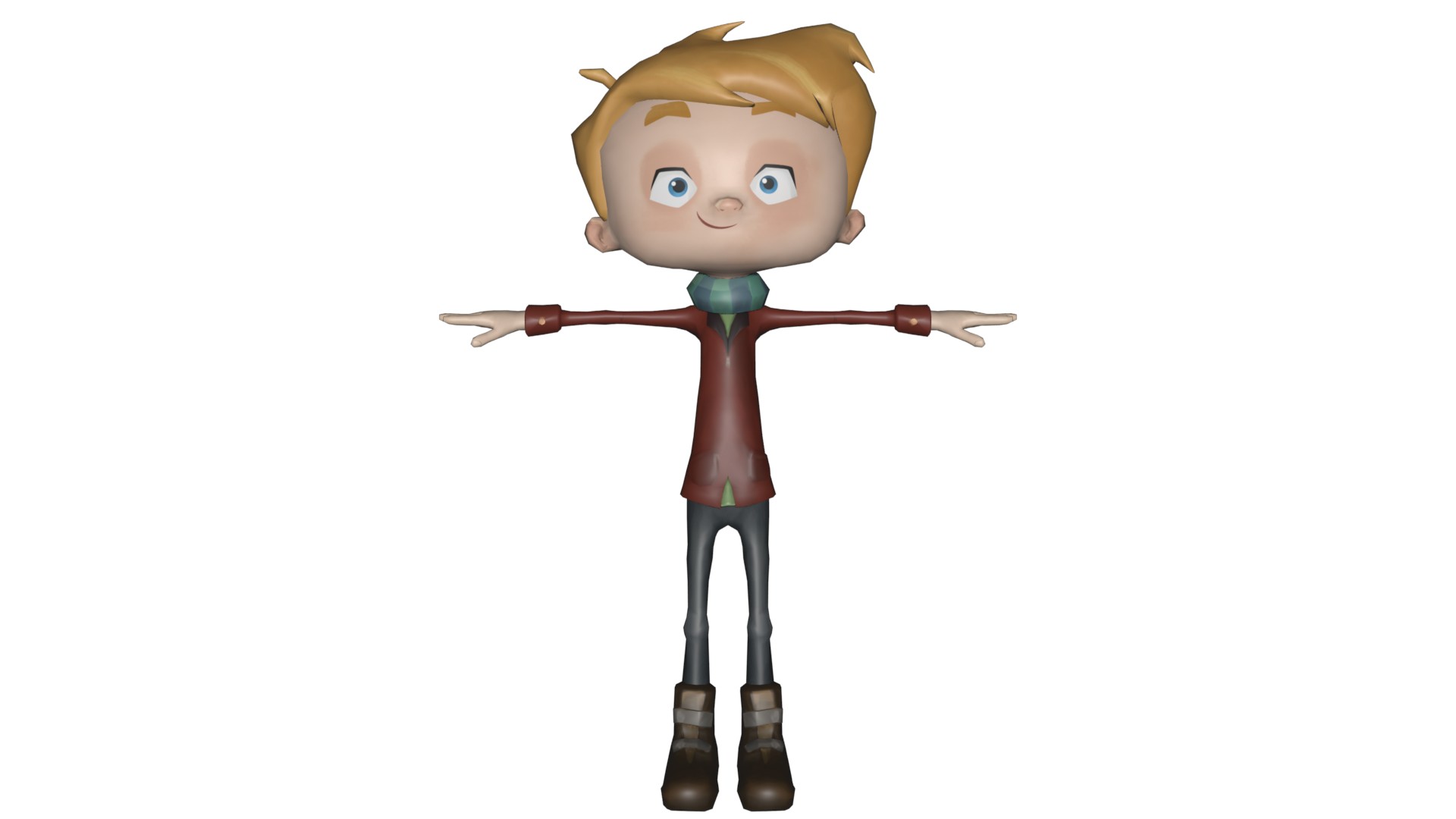}\label{fig:humanoid-boy}}
\subfloat[Girl humanoid model.]{\includegraphics[width=0.4\textwidth]{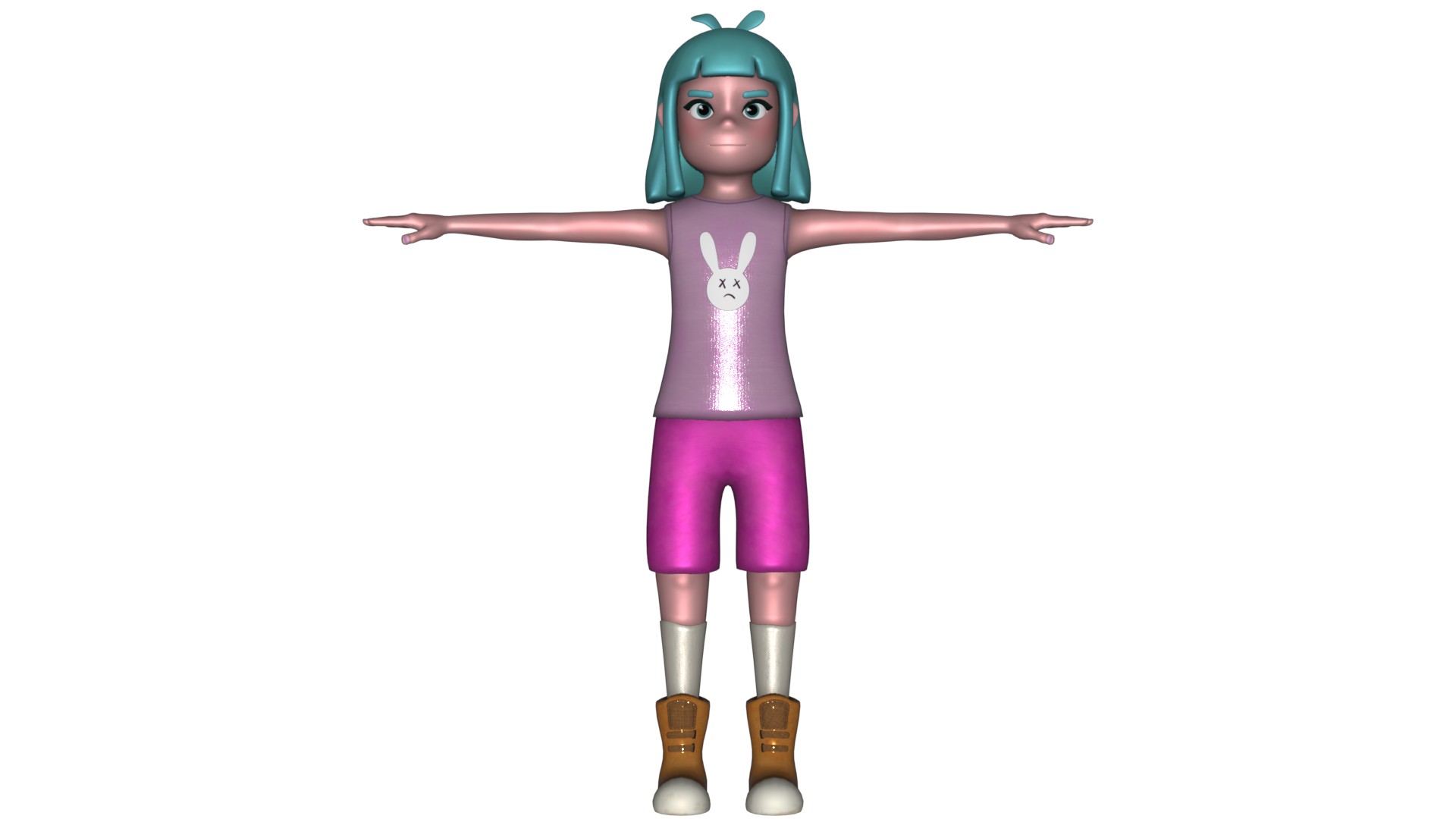}\label{fig:humanoid-girl}}\hskip1ex
\subfloat[Mousy humanoid model.]{\includegraphics[width=0.4\textwidth]{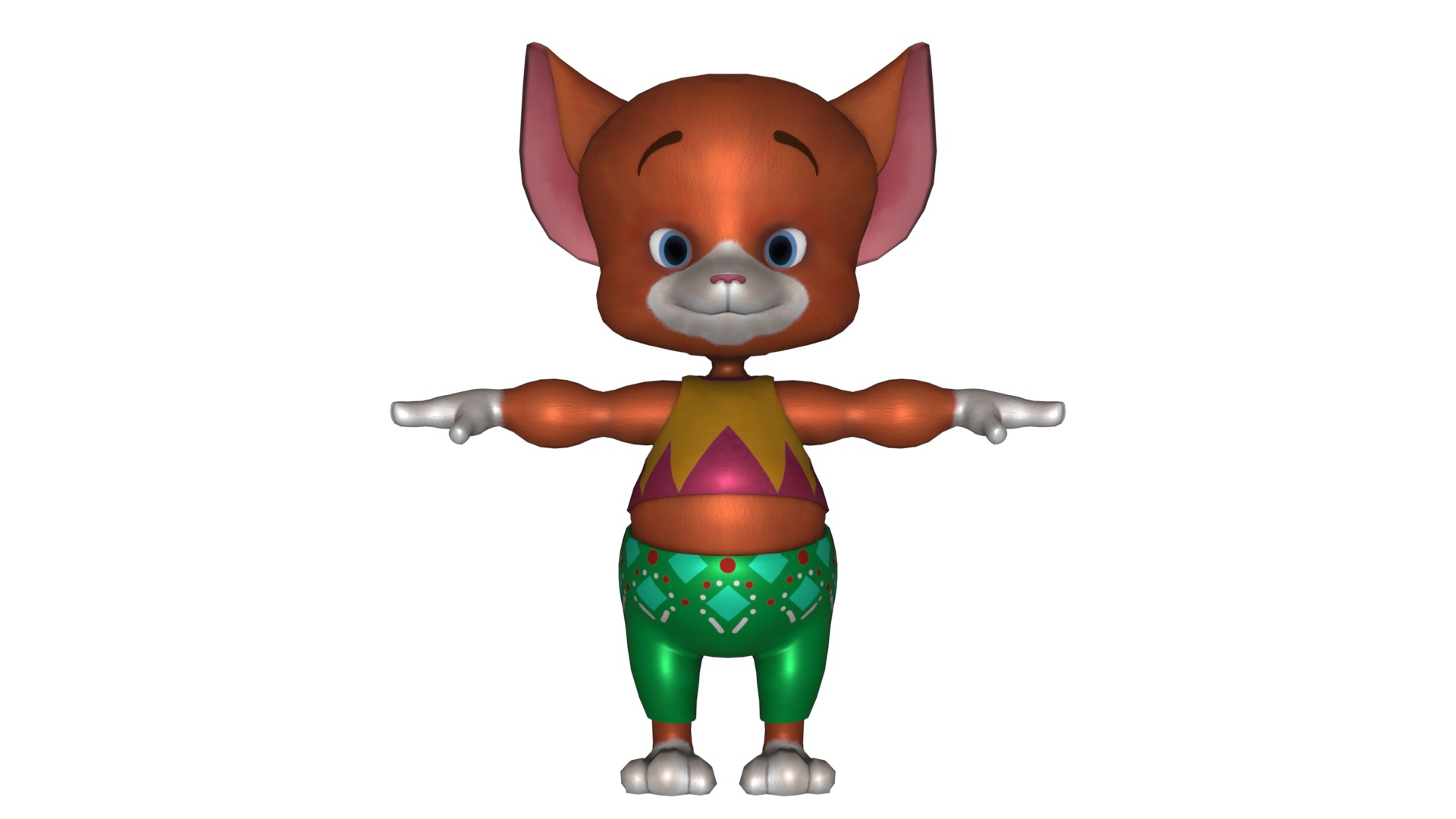}\label{fig:humanoid-mousy}}
\subfloat[Robot humanoid model.]{\includegraphics[width=0.4\textwidth]{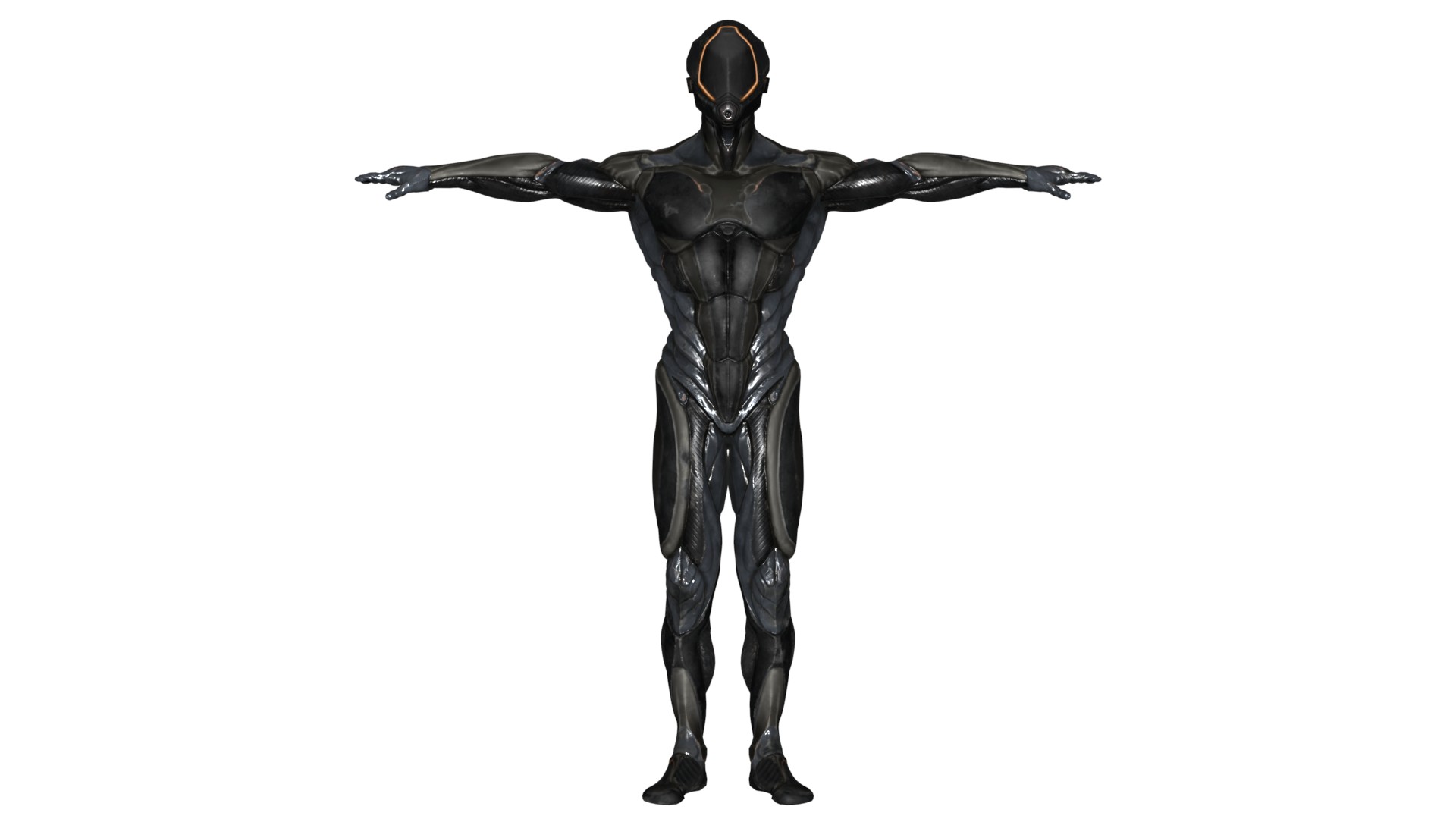}\label{fig:humanoid-robot}}
\caption{Showcase of available humanoid models: (a) Boy, (b) Girl, (c) Mousy, and (d) Robot, each offering unique characteristics for dance performances.}
\label{fig:humanoids}
\end{figure}

Users can explore these AI-choreographed movements through different humanoid models, including Boy, Girl, Mousy, and Robot (Fig. \ref{fig:humanoids}), offering a rich and diverse visualization of the generated dance sequences and allowing them to tailor their dance performance to their preferences.

In additionally, the platform also enables users to import their personal dance videos. This feature allows users to compare their own movements with those produced by the AI. This comparison can inspire and motivate users to engage more actively in dancing, fostering a deeper connection between the user and the art of dance.

\subsubsection{Findings on Music to Dance Generation}

Generating realistic and engaging music-conditioned dance sequences is a complex challenge, which requires addressing two crucial aspects. The first is the temporal challenge, where the generated dance sequences should synchronize with the given music's beat and rhythm while maintaining naturalness. Equally important is the spatial challenge, which demands that the generated dance moves be both physically realistic and aesthetically pleasing. 

To address these challenges, various methods have been developed, which can be divided into two main categories: autoregressive and non-autoregressive approaches. Autoregressive methods \cite{sun2022you, huang2020dance, yang2023longdancediff} generate future motions based on the previous ones, effectively capturing the flow of the dance over time. However, they often struggle to produce long-term sequences, resulting in a freezing effect where movements become repetitive and static. On the other hand, non-autoregressive methods \cite{li2024lodge, chen2021choreomaster, tseng2023edge} handle each motion segment independently and use simple temporal smoothing techniques to smooth out the transitions between segments, but can lead to abrupt and unnatural transitions between poses.

% comment from reviewer: The paper could benefit from more extensive technical details about the underlying AI models, especially how diffusion models are applied specifically to music-driven dance generation.

To overcome these limitations, we utilize DanceFusion \cite{mypaper}, a conditional diffusion model designed for generating realistic dance movements that are conditioned on music as well as past and future motions. Particularly, the diffusion process is represented as a Markov chain process, gradually adding noise to real data. At each step of this process, Gaussian noise is injected, with the amount of noise controlled by a predefined schedule, $\{a_t \in (0, 1)\}_{t=1}^T$. With latent $\{z_t\}_{t=0}^T$ that follow a forward noising process $q(z_t|x)$, where $x\sim p(x)$ is drawn from the data distribution, the forward noising process is mathematically defined as:
\begin{equation}\label{eq:4.1}
q(z_t | x) \sim N(z_t; \sqrt{a_t}x, (1 - a_t)I).
\end{equation}

In the setting with music conditioning $m$, past motions $x_p$ and future motion $x_f$, the author reverse the forward diffusion process by learning to estimate \( \hat{x}_\theta (z_t, t, m, x_p, x_f) \approx x \) with model parameters \( \theta \) for all \( t \). They optimize \( \theta \) with the "simple" objective introduced in Ho \etal \cite{ho2020denoising}:
\[
\mathcal{L}_{\text{simple}} = \mathbb{E}_{x, t} \left[ \left\| x - \hat{x}_\theta (x_t, t, m, x_p, x_f) \right\|_2^2 \right].
\]

DanceFusion effectively combines the strengths of autoregressive and non-autoregressive techniques, allowing for the creation of arbitrarily long performances. The authors of this model have also demonstrated its effectiveness in enhancing the quality of generated dance sequences. This capability makes DanceFusion suitable for applications involving diverse and spontaneous music selections. By leveraging DanceFusion, we aim to generate high-quality dance sequences that are both temporally and spatially coherent, providing a more engaging and realistic experience for users.

\subsubsection{Motion Representation}

\begin{figure}[h!]
    \centering
    \includegraphics[width=0.45\linewidth]{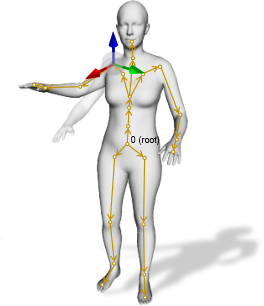}
    \caption{Human pose and shape representation in rotation-based.}
    \label{fig:human-pose}
\end{figure}

% To effectively capture the complexities of human movement, suitable motion representation method should be employed. Various parameterizations, such as Euler angles, axis angles, or quaternions, can be used to represent the rotations of poses in 3D space. However, these methods may not fully capture the intricate shapes and subtle deformations of the human body during movement.

% To address this limitation, we investigate the use of statistical mesh models and adopt the SMPL model \cite{loper2023smpl, pavlakos2019expressive} for our system.This decision was motivated by several factors. Firstly, the SMPL model aligns perfectly with the output of DanceFusion, our chosen dance generation model. Secondly, the SMPL model provides a more detailed representation of the human form, capturing not only the position and orientation of body parts but also the underlying shape and surface details, as shown in Fig. \ref{fig:human-pose}. 

% The SMPL model is controlled by two key sets of parameters: pose and shape. These parameters work together to generate a 3D mesh representation of a human figure in a specific pose and with a distinct shape. Pose parameters, denoted by $\theta$, define the relative orientation of each joint within a hierarchical skeletal structure. This structure consists of 24 joints, with each joint's orientation determined by its rotation relative to its parent joint.

To capture the complexities of human movement, we need an effective motion representation. Common parameterizations like Euler angles, axis angles, and quaternions describe 3D rotations but may miss subtle body deformations during movement.

To overcome this, we use statistical mesh models and adopt the SMPL model \cite{loper2023smpl, pavlakos2019expressive}, chosen for its compatibility with DanceFusion, our dance generation model, and its detailed representation of human form, capturing body position, orientation, and surface details, as shown in Fig. \ref{fig:human-pose}.

The SMPL model relies on two parameter sets: pose and shape. Pose parameters ($\theta$) define each joint's orientation within a 24-joint hierarchy, where each joint’s rotation is relative to its parent.

\subsection{Implementation}

\subsubsection{Motion Importing}

The process of importing and visualizing the dance files generated by DanceFusion within the Blender environment is a crucial step. To facilitate this process, we utilize a specialized Blender add-on known as ``SMPL to FBX,''\footnote{\url{https://github.com/softcat477/SMPL-to-FBX}}. This add-on enables us to seamlessly import the SMPL model files generated by DanceFusion into Blender, where we can further refine and visualize the dance movements. 

After refining the dance sequences in Blender, we export the files in FBX format, which is a widely supported format for 3D models and animations. Finally, we import the FBX files into Unity for visualization and integration with our application DanceDuo.

\subsubsection{Motion Estimation from Video}

To estimate the SMPL pose from video, we employ the ROMP model, a state-of-the-art, one-stage method for real-time, monocular, multi-person 3D mesh recovery \cite{sun2021monocular}. ROMP's streamlined architecture enables efficient estimation of SMPL 3D pose parameters from a single image, making it an ideal choice for our application.

To optimize computational efficiency and processing speed, we adopt a selective frame extraction approach. Instead of extracting SMPL parameters from every frame, we extract them at regular intervals and interpolate the intermediate frames. This strategy significantly reduces the computational load while maintaining the required processing speed for our application.

%review
Let $\theta_{j,f}$ denotes the rotation of joint $j$ at frame $f$. To estimate the rotation of a joint at frame $f$, we first estimate the pose at the preceding frame $f - 1$ and the succeeding frame $f + 1$, obtaining $\theta_{j, f - 1}$ and $\theta_{j, f + 1}$, respectively. We then calculate the rotation of joint $j$ at frame $f$ using the average of the two estimated rotations:

\begin{equation}
    \theta_{j, f} = \frac{\theta_{j, f - 1} + \theta_{j, f + 1}}{2}.
\end{equation}

By applying this interpolation technique to all 24 joints, we obtain the interpolated pose at frame $f$. In cases where $f + 1$ exceeds the total number of frames, we estimate the pose at frame $f$ using the available information.

\subsubsection{Scoring System}

% For the scoring system in DanceDuo, we aimed for more than simple similarity calculations, focusing on user engagement and motivation regardless of dance skill level. We designed a formula that scores between $6$ and $8$ out of $10$, helping users feel accomplished even while learning.

% Our approach transforms the raw "similarity score" into a bell-shaped distribution, like a normal curve. This keeps most scores in a balanced range, avoiding extremes like $0$ or $1$, which could be discouraging.

For the scoring system in DanceDuo, we aimed for more than simple similarity calculations, focusing on user engagement and motivation regardless of dance skill level. We designed a formula that scores between $6$ and $8$ out of $10$, helping users feel accomplished even while learning.

Our approach transforms the raw "similarity score" into a beta distribution, and making it like a normal curve. This keeps most scores in a balanced range, avoiding extremes like $0$ or $1$, which could be discouraging.

We calculate accuracy based on two primary metrics:

\begin{enumerate}
    \item \textbf{Mean Per Joint Position Error (MPJPE):} MPJPE measures the average distance between corresponding joints in predicted and ground truth poses. Given a set of predicted joint positions $P$ and ground truth joint positions $G$, the MPJPE is calculated as follows:
    \begin{equation}
        \text{MPJPE}(P, G) = \frac{1}{N} \sum_{i=1}^{N} \| P_i - G_i \|,
    \end{equation}
    where $ N $ is the number of joints, and $ \| P_i - G_i \| $ denotes the Euclidean distance between the $ i $-th joint's predicted position $ P_i $ and the ground truth position $ G_i $.

    \item \textbf{Mean Per Joint Angle Error (MPJAE):} MPJAE measures the average angular error between corresponding joints' orientations in predicted and ground truth poses. Given the predicted joint angles $ A_p $ and ground truth joint angles $ A_g $, the MPJAE is calculated as follows:
    \begin{equation}
        \text{MPJAE}(A_p, A_g) = \frac{1}{N} \sum_{i=1}^{N} \theta(A_{p_i}, A_{g_i}),
    \end{equation}
    where $ N $ is the number of joints, and $ \theta(A_{p_i}, A_{g_i}) $ represents the angular difference between the $ i $-th joint angle's predicted orientation $ A_{p_i} $ and the ground truth orientation $ A_{g_i} $. The angular difference $ \theta $ can be computed using the dot product of unit vectors representing the joint angles:
    \begin{equation}
        \theta(A_{p_i}, A_{g_i}) = \cos^{-1} \left( \frac{A_{p_i} \cdot A_{g_i}}{\|A_{p_i}\| \|A_{g_i}\|} \right).
    \end{equation}
\end{enumerate}

% \begin{figure}[t!]
% \centerline{\includegraphics[width=0.7\textwidth]{figure/beta ppf.jpg}}
% \caption{Percentage Point Function of the beta distribution with parameters $\alpha = 3$ and $\beta = 1$. This function transforms raw accuracy scores into refined similarity scores.}
% \label{fig:beta_ppf}
% \end{figure}

% Next, we map the combined similarity score to a Beta distribution using the Percentage Point Function (PPF). Our choice of the Beta distribution was motivated by two key factors. Firstly, its shape can be flexibly controlled by the parameters $\alpha$ and $\beta$ enabling us to create a bell-shaped curve that closely resembles the normal distribution. %, as illustrated in Fig. \ref{fig:beta}. 
% This flexibility allows us to fine-tune the distribution to best suit our scoring needs. Secondly, the Beta distribution has the advantageous property of mapping values back to the $[0, 1]$ range, which is ideal for our scoring system. For our application, we chose $\alpha = 3$ and $\beta = 1$ to achieve a well-balanced bell curve. %Fig. \ref{fig:beta_ppf} illustrates the Percentage Point Function of this beta distribution. %This characteristic ensures that our final scores remain within a consistent and interpretable range, regardless of the input values. For our application, we chose $\alpha = 3$ and $\beta = 1$ to achieve a well-balanced bell curve. Fig. \ref{fig:beta_ppf} illustrates the Percentage Point Function of this beta distribution, which plays a crucial role in our scoring mechanism.

We map the combined similarity score to a Beta distribution using the Percentage Point Function (PPF). The Beta distribution offers two key benefits: flexibility in shaping the curve with parameters $\alpha$ and $\beta$, allowing us to create a bell-shaped curve similar to the normal distribution, and its ability to map values to the $[0, 1]$ range, ideal for scoring. We set $\alpha = 3$ and $\beta = 1$ to achieve a well-balanced bell curve.

% \begin{figure}[t!]
% \centerline{\includegraphics[width=0.7\textwidth]{figure/beta_distribution.png}}
% \caption{Probability Density Functions (PDFs) of the Beta distribution for different parameter values. The Beta distribution can be adjusted to create a bell-shaped curve within the $[0, 1]$ range with $\alpha$ and $\beta$. The figure is from \cite{wikipediaBetaDistribution}.}
% \label{fig:beta}
% \end{figure}

To calculate the accuracy at each frame, we consider both MPJPE and MPJAE. For each frame $f$, we compute the accuracy by averaging MPJPE and MPJAE:
\begin{equation}
    \text{accuracy}_f = 1 - \frac{\text{MPJPE}_f + \text{MPJAE}_f}{2}.
\end{equation}

The raw accuracy score is then refined using the Beta PPF:
\begin{equation}
    \text{score}_f = \text{PPF}(\text{accuracy}_f).
\end{equation}

Finally, the total score for a sequence is the average of all frame scores, scaled to a 100-point system.

\section{User Study}

\subsection{Setup}

% Five participants were invited to interact with the application and subsequently complete a detailed survey designed to gather in-depth feedback on their user experience. The survey featured open-ended questions focusing on usability, interface design, and feature effectiveness. Participants were encouraged to provide honest feedback, sharing both positive impressions and areas for improvement.

Five participants were invited to interact with the application and complete a survey on their user experience. The survey featured open-ended questions on usability, interface design, and feature effectiveness.

\begin{figure}[t!]
    \centering
    \includegraphics[width=0.75\textwidth]{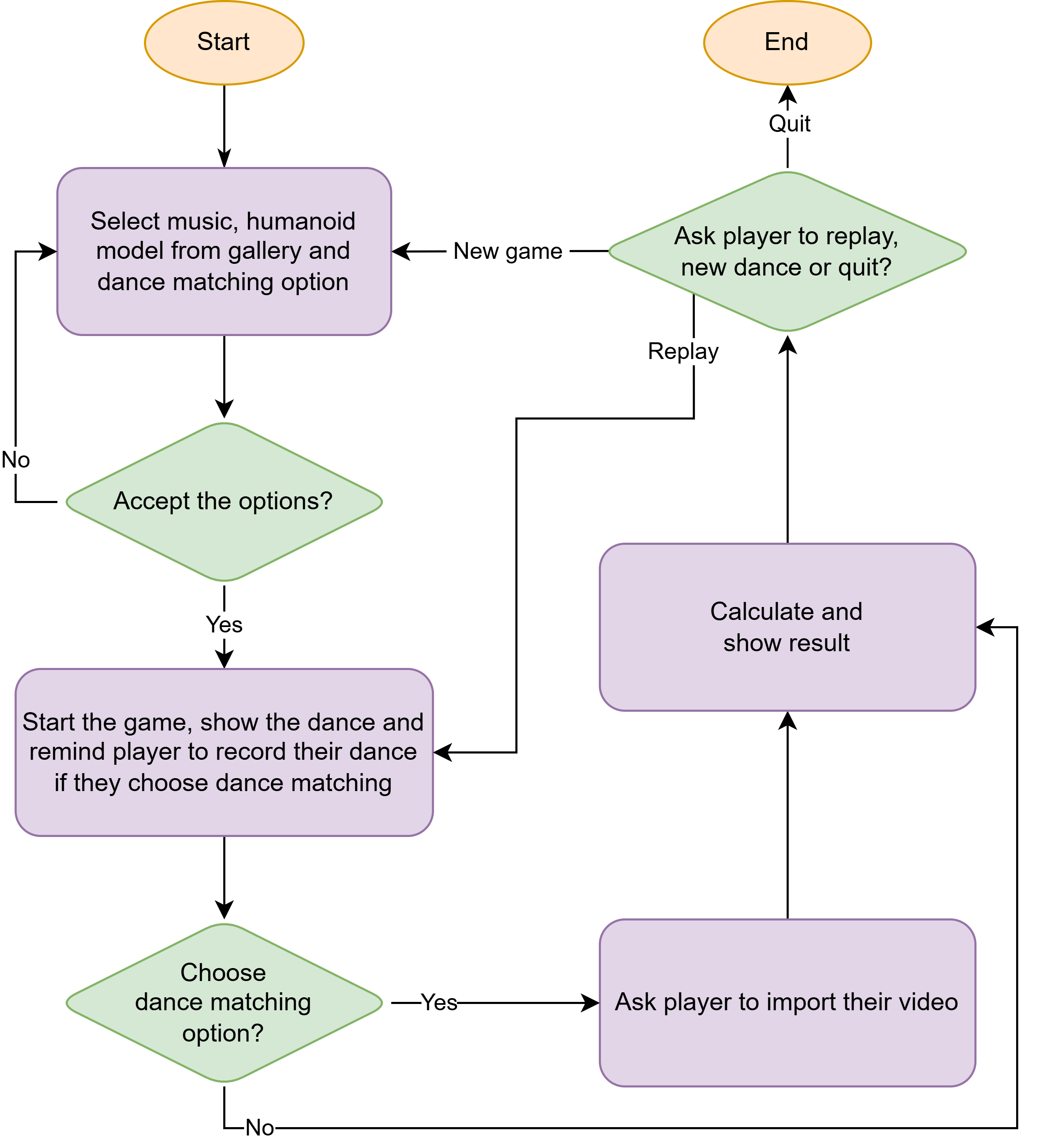}
    \caption{Flowchart illustrating the user’s process through the application}
    \label{fig:workflow}
\end{figure}

\subsection{User Workflow}

% The application was designed with a user-centric approach to provide an intuitive experience. Fig. \ref{fig:workflow} shows the user journey, detailing each step from launch to final interaction options.

The application follows a user-centric design for an intuitive experience, as illustrated in Fig. \ref{fig:workflow}, showing the journey from launch to final options.

\begin{figure}[t!]
\centering
\subfloat[UI to select musics]{\includegraphics[width=0.7\textwidth]{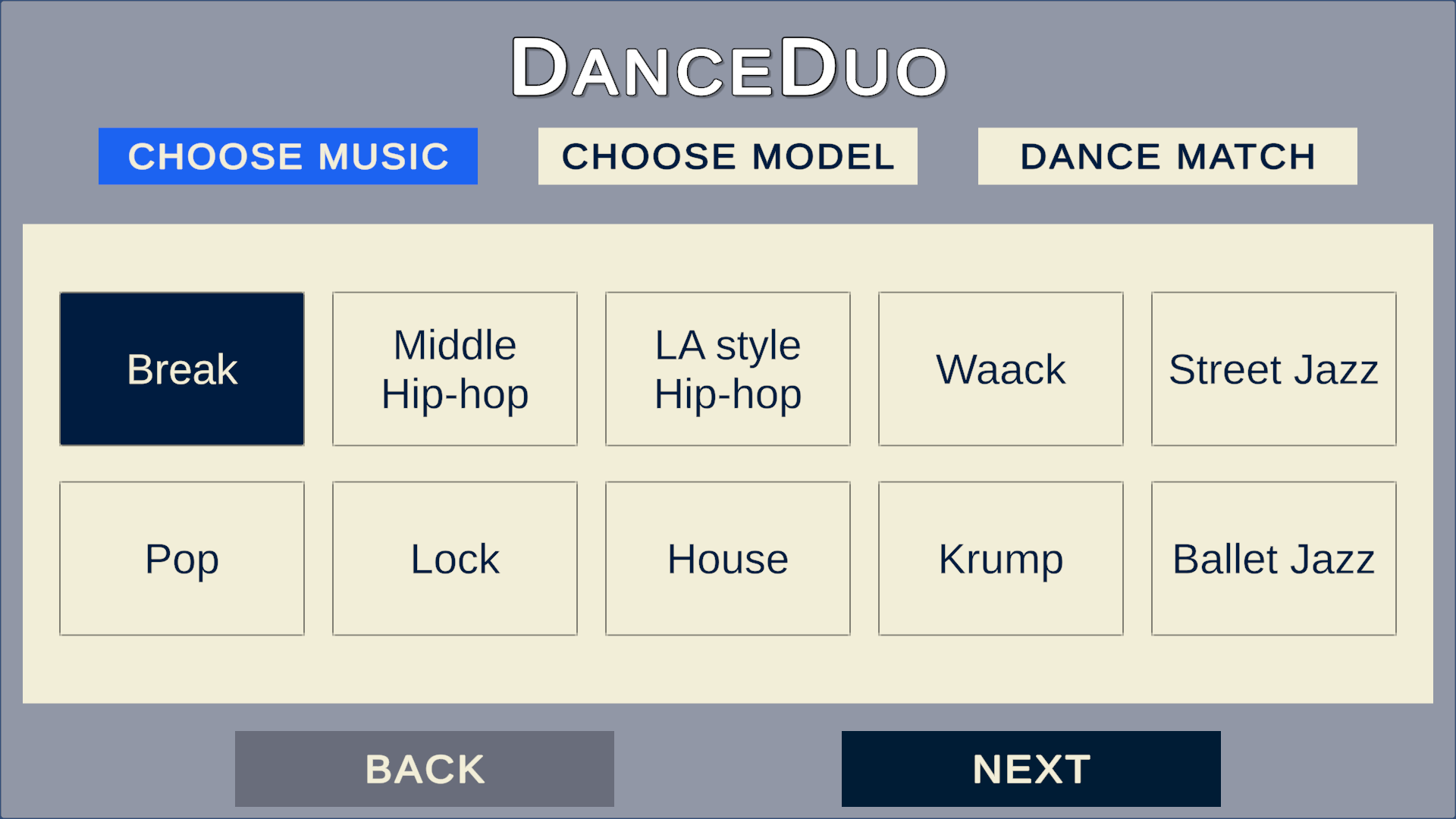}\label{fig:ui-music}} \\
\subfloat[UI to select humanoid models]{\includegraphics[width=0.7\textwidth]{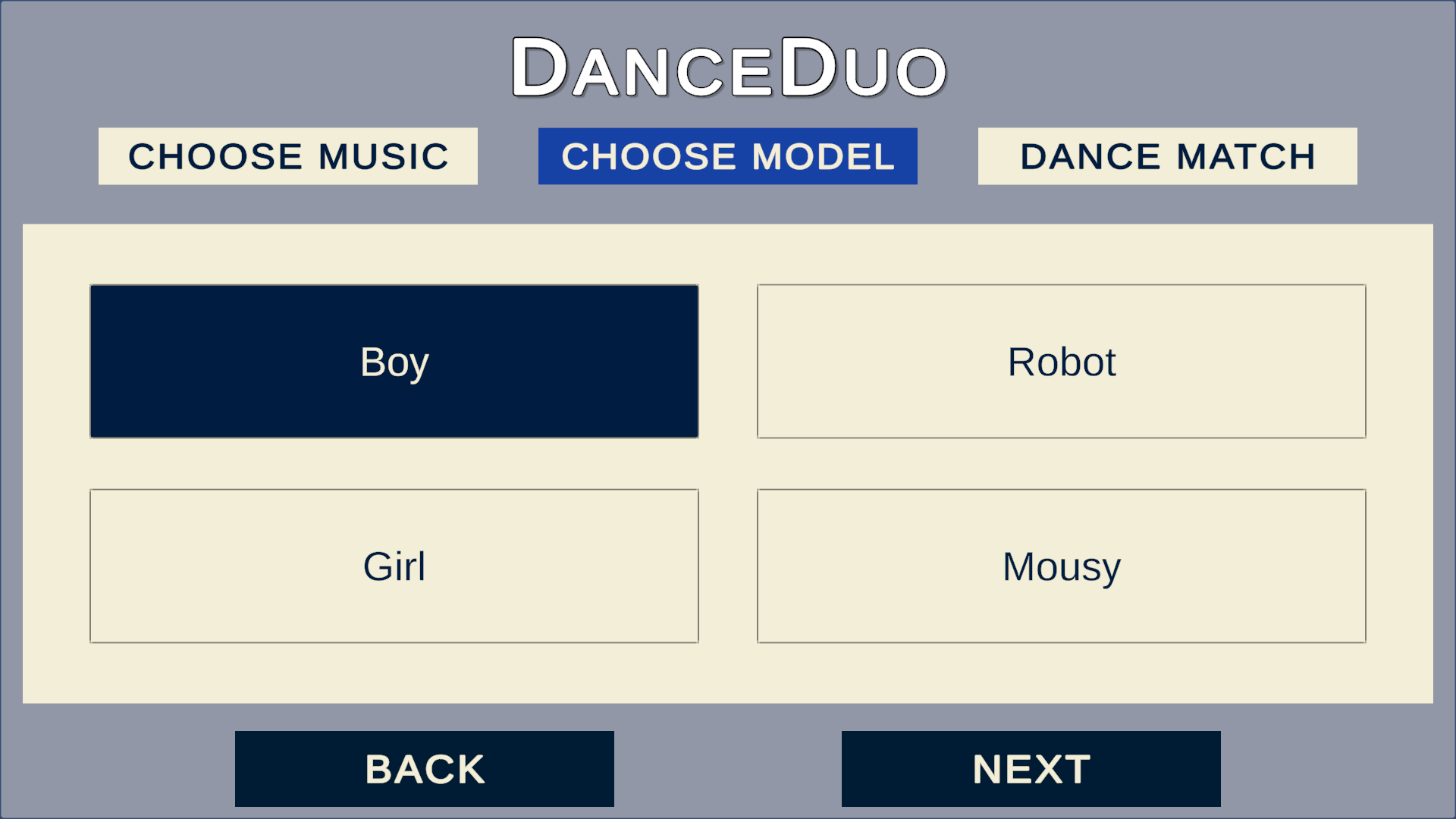}\label{fig:ui-humanoid}}
\caption{UI where user choose music and humanoid model to showcase the dance.}
\label{fig:selecting-music-and-humanoid}
\end{figure}

\begin{figure}[t!]
\centering
\subfloat[Section where user watch the generated dance performance with selected music and humanoid model.]{\includegraphics[width=0.7\textwidth]{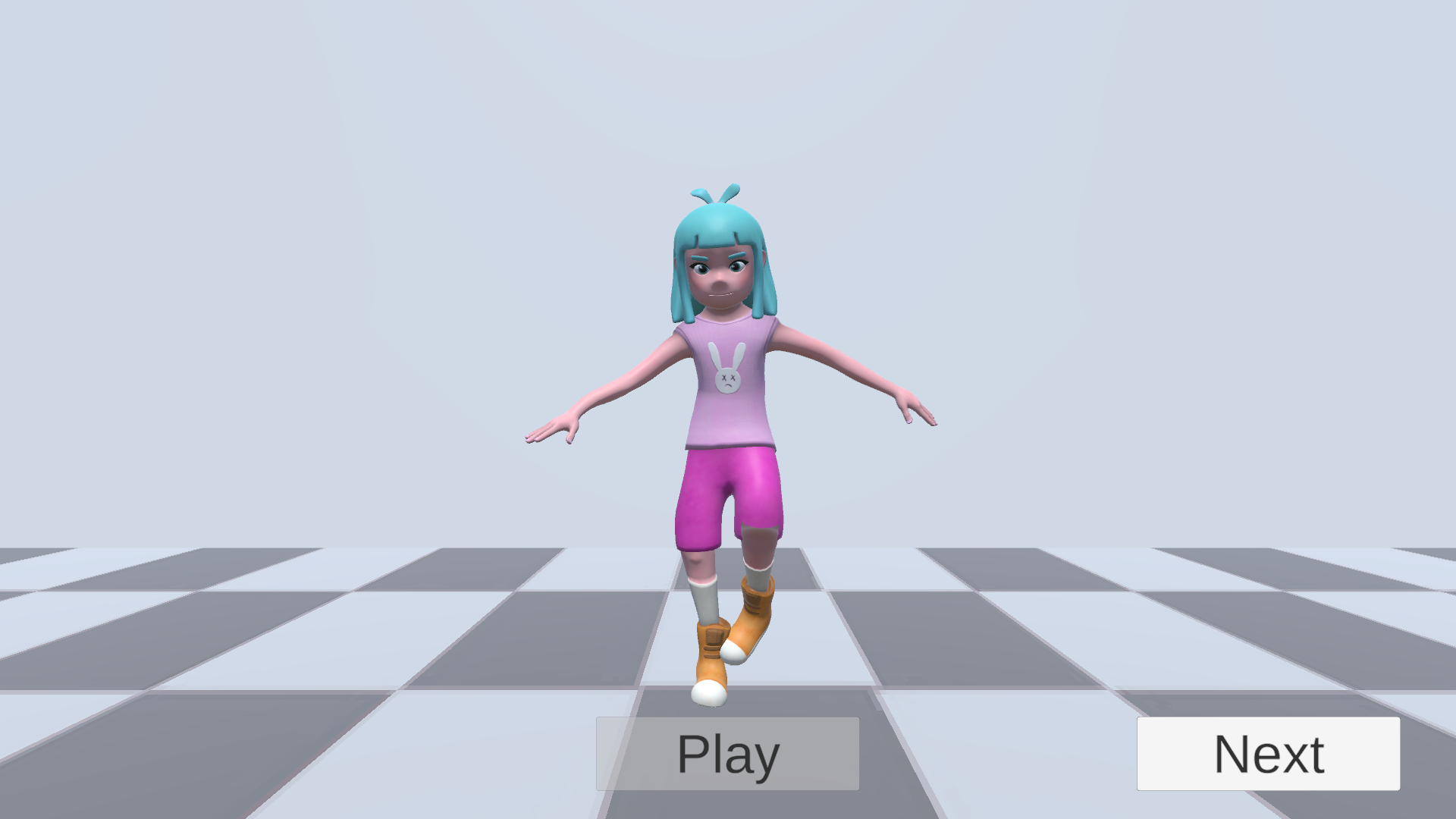}\label{fig:ui-show-dance}} \\
\subfloat[UI to select humanoid models]{\includegraphics[width=0.7\textwidth]{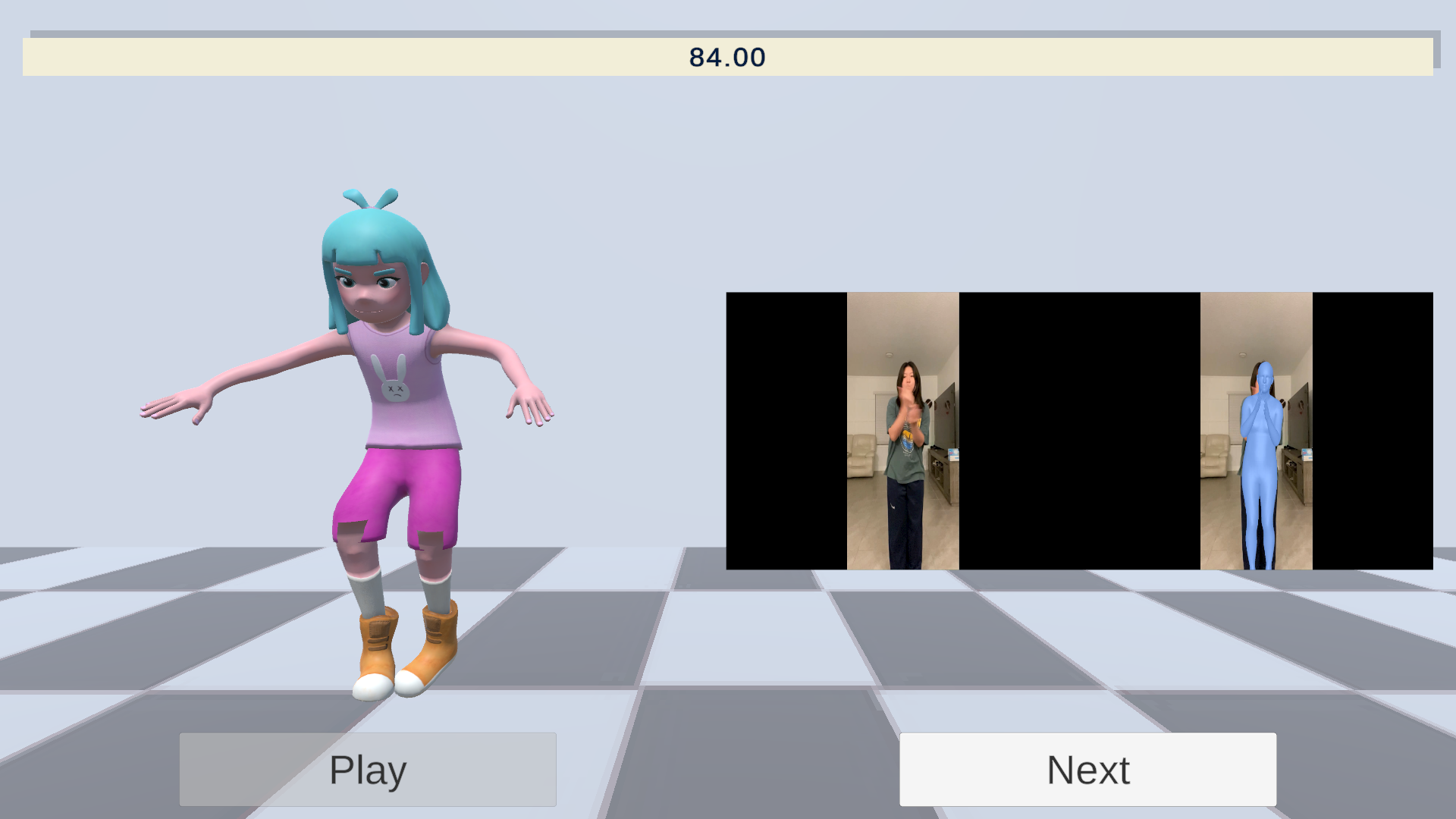}\label{fig:ui-show-score-with-vid}}
\caption{User can watch the dance performance again along side with their dance video. They can also see the score at the top of the screen.}
\label{fig:show-score-and-show-dance}
\end{figure}

Users start by selecting a music track and a humanoid model in a gallery interface, as shown in Fig. \ref{fig:selecting-music-and-humanoid}. They then watch the generated dance performance, as shown in Fig. \ref{fig:ui-show-dance} and can record and upload their own dance video. The app processes this video, compares it to the AI-generated dance, and provides a similarity score. Users can view their score and watch the AI and their own videos side-by-side, as shown in Fig. \ref{fig:ui-show-score-with-vid}.

\subsection{Results}

The user study provided valuable insights into the application's strengths and potential areas for improvement:

\paragraph{\textbf{Usability and Interface Design}}

%Participants generally found the application intuitive and user-friendly. The interface design and user flow received positive feedback, with users appreciating the seamless navigation across different sections. The simplicity of the design was highlighted as a key strength, contributing to an overall smooth user experience. However, some participants suggested improvements. A few users expressed the need for more detailed instructions or context-sensitive help, particularly regarding the functionality of certain buttons. While the simplicity of the application was appreciated, this feedback suggests that additional guidance could further enhance user understanding and interaction.

Participants generally found the application intuitive and user-friendly, praising its simple design and seamless navigation. However, some users suggested providing more detailed instructions or context-sensitive help to enhance user understanding and interaction.

\paragraph{\textbf{Interactive Features}}

%The dance comparison feature was consistently identified as one of the most engaging aspects of the application. Participants praised its interactive nature and motivational impact, highlighting the value of comparing their own dance performances with AI-generated dances. This feature was seen as both entertaining and practical, offering users a tangible way to assess and improve their dance skills. However, some users suggested that the feature could be improved by providing more specific, actionable feedback on how to refine their dance moves.

The dance comparison feature was consistently identified as one of the most engaging aspects of the application. Users praised its interactive nature and motivational impact, highlighting the value of comparing their own dance performances with AI-generated dances. Some users suggested that providing more specific, actionable feedback could further improve the feature.

\paragraph{\textbf{Variety and User Engagement}}

%Participants expressed strong appreciation for the diverse options available within the application. The wide array of music genres and humanoid models contributed to a highly engaging and personalized experience. Users found the ability to customize these elements both entertaining and meaningful, allowing them to tailor their interactions according to their preferences. This variety was identified as a key factor in enhancing user engagement and enjoyment.

Participants appreciated the diverse options available within the application, particularly the wide array of music genres and humanoid models. Users found the ability to customize these elements entertaining and meaningful, enhancing their engagement and enjoyment.

\paragraph{\textbf{Output Quality}}

Feedback on the quality of the generated dance animations was generally positive, with users satisfied with the fluidity and the visual representation of the movements. However, some participants pointed out areas for improvement, particularly regarding synchronization between the music and the generated dance movements. Refining this aspect could further enhance the overall quality of the output.

% \section{Limitations and Future Works}

% There are limitations in the current version of the DanceDuo that should be addressed in future updates. A primary limitation is the absence of real-time features, especially during the processing of user-imported dance videos. Currently, users experience delays as they wait for the motion estimation process to complete, which can disrupt the overall flow of interaction. To improve this, future updates can focus on implementing real-time processing capabilities, offering a smoother and more immediate user experience.

% Another limitation is the restricted audio selection, as users are limited to a predefined set of tracks. The ability to import and use custom audio will be a key feature in future versions, allowing for a more personalized and dynamic interaction between user-selected music and AI-generated dance movements.

% Additionally, participants identified a need for more detailed instructions and clearer guidance within the application. Many users suggested the inclusion of descriptive labels for buttons and hover-over tooltips to provide contextual information on various features. By addressing these usability concerns, future updates will aim to enhance the intuitiveness of the interface, making navigation and feature exploration more seamless and user-friendly.

\section{Conclusion}

In this paper, we introduced DanceDuo, a novel platform designed to generate in-the-wild, AI-driven dance sequences synchronized with music tracks, using diffusion models. By integrating a diverse selection of dance styles and humanoid models, DanceDuo enables users to explore and engage with AI-generated choreography in a personalized and interactive manner. The platform’s ability to import user-generated dance videos for comparison with AI-choreographed movements further enhances user engagement, promoting learning, creativity, and skill improvement.

% Our system leverages the DanceFusion model to address both the temporal and spatial challenges of music-conditioned dance generation, producing coherent and natural dance sequences that maintain long-term synchronization with the accompanying music.

The user study demonstrated that DanceDuo provides an intuitive and engaging experience, particularly highlighting the effectiveness of its dance comparison feature. While the feedback was generally positive, it also revealed several areas for improvement, including the need for real-time features, enhanced usability guidance, and support for custom audio tracks.

In future work, we aim to address these limitations by integrating real-time processing capabilities and expanding the platform’s functionality to include more customizable features. These enhancements will further improve the user experience, encouraging deeper engagement and creativity in the intersection of music, dance, and AI.

% ----------

% There are limitations in the current version of the DanceDuo that should be addressed in future updates. A primary limitation is the absence of real-time features, especially during the processing of user-imported dance videos. Currently, users experience delays as they wait for the motion estimation process to complete, which can disrupt the overall flow of interaction. To improve this, future updates can focus on implementing real-time processing capabilities, offering a smoother and more immediate user experience.

% Another limitation is the restricted audio selection, as users are limited to a predefined set of tracks. The ability to import and use custom audio will be a key feature in future versions, allowing for a more personalized and dynamic interaction between user-selected music and AI-generated dance movements.

% Additionally, participants identified a need for more detailed instructions and clearer guidance within the application. Many users suggested the inclusion of descriptive labels for buttons and hover-over tooltips to provide contextual information on various features. By addressing these usability concerns, future updates will aim to enhance the intuitiveness of the interface, making navigation and feature exploration more seamless and user-friendly.

\section*{Acknowledgment}

This research is funded by University of Science, VNU-HCM, under grant number CNTT 2024-16.

%
% ---- Bibliography ----
%
% BibTeX users should specify bibliography style 'splncs04'.
% References will then be sorted and formatted in the correct style.
%
\bibliographystyle{splncs04}
\bibliography{main}

\end{document}